# The Origin of Element Abundance Variations in Solar Energetic Particles

**Donald V. Reames**

Institute for Physical Science and Technology, University of Maryland, College Park, MD 20742-2431 USA, email: dvreames@umd.edu

**Abstract** Abundance enhancements, during acceleration and transport in both gradual and impulsive solar energetic particle (SEP) events, vary approximately as power laws in the mass-to-charge ratio [$A/Q$] of the ions. Since the $Q$-values depend upon the electron temperature of the source plasma, this has allowed a determination of this temperature from the pattern of element-abundance enhancements and a verification of the expected inverse-time dependence of the power of $A/Q$ for diffusive transport of ions from the SEP events, with scattering mean free paths found to be between 0.2 and 1 AU. SEP events derived from plasma of different temperatures map into different regions in typical cross-plots of abundances, spreading the distributions. In comparisons of SEP events with temperatures above 2 MK, impulsive events show much broader non-thermal variation of abundances than do gradual events. The extensive shock waves accelerating ions in gradual events may average over much of an active region where numerous but smaller magnetic reconnections, "nanojets", produce suprathermal seed ions, thus averaging over varying abundances, while an impulsive SEP event only samples one local region of abundance variations. Evidence for a reference He/O abundance ratio of 91, rather than 57, is also found for the hotter plasma. However, while this is similar to the solar-wind abundance of He/O, the solar-wind abundances otherwise provide an unacceptably poor reference for the SEP abundance enhancements, generating extremely large errors.

# 1. Introduction

The relative abundances of the chemical elements and their variation in time have been a major factor in distinguishing the particle sources and in understanding the physical processes of their acceleration and transport (Reames, 1999). For solar energetic particle (SEP) events, two distinct mechanisms of particle acceleration have been identified that dominate the SEP events we call "impulsive" and "gradual" (see reviews by Gosling, 1993; Lee, 1997; Reames, 1999, 2013, 2015; Desai *et al.*, 2004; Mason, 2007; Kahler, 2007; Cliver and Ling, 2007, 2009; Kahler *et al.*, 2012; Wang *et al.*, 2012). Gradual or long-duration events, the most intense and energetic of the SEP events, involve particle acceleration at shock waves driven out from the Sun by coronal mass ejections (CMEs; see Kahler *et al.*, 1984; Mason, Gloeckler, and Hovestadt, 1984; Kahler, 1992, 1994, 2001; Gopalswamy *et al.*, 2004, 2012; Cliver, Kahler, and Reames, 2004; Rouillard *et al.*, 2011, 2012; Lee, Mewaldt, and Giacalone, 2012). Element abundances in these events show significant variation with time during an event and from one event to the next. The variations can be organized as an increasing or decreasing power-law function of the mass-to-charge ratio [$A/Q$] of the elements (Breneman and Stone, 1985) but these abundances, averaged over many events, are a measure of the abundances of the solar corona (Meyer, 1985; Reames, 1995, 2014). It has recently been shown that the abundance variations result from varying source plasma temperatures that determine $Q$ and are magnified by the $A/Q$ dependence of transport from the shock (Reames, 2016).

Impulsive SEP events are smaller events with shorter durations characterized by unusual abundances, first noticed as the little $^3$He-rich events with $^3$He/$^4$He, often >1, exceeding the abundance in the solar wind by factors of up to $10^4$ (Hsieh and Simpson, 1970; Serlemitsos and Balasubrahmanyan, 1975). Subsequent measurements found ten-fold enhancement in Fe/O, uncorrelated with $^3$He/$^4$He (Mason *et al.*, 1986), then 1000-fold enhancements of heavy elements ($Z$ >50)/O (Reames, 2000; Mason *et al.*, 2004; Reames and Ng, 2004). The $^3$He-rich events are associated with streaming electrons and Type-III radio bursts (Reames, von Rosenvinge, and Lin, 1985; Reames and Stone, 1986) which may excite waves that resonate with $^3$He (Temerin and Roth, 1992; Roth and Temerin, 1997). The strong $A/Q$-dependent enhancements may be produced by Fermi acceleration in islands of magnetic reconnection (Drake *et al.*, 2009) in solar flares and jets (Shimojo and Shibata, 2002; see also Kahler, Reames, and Cliver. 2015). The pattern of $Q$-values in the $A/Q$-dependence of the strong abundance enhancements provides best-fit source plasma temperatures of 2–4 MK (Reames, Cliver, and Kahler. 2014a, b, 2015). This work followed earlier attempts to link abundances with temperature (Reames, Meyer, and von Rosenvinge, 1994; Cohen *et al.*, 1999)

The evidence of the separation of the two acceleration mechanisms is complicated by the fact that shock waves can reaccelerate residual ions from earlier impulsive SEP events (Mason, Mazur, and Dwyer, 1999; Tylka *et al.*, 2001; Desai *et al.*, 2003). In fact, the shock may preferentially select seed particles from impulsive suprathermal ions when they are present (Tylka *et al.*, 2005; Tylka and Lee, 2006), causing complex energy dependence in abundance ratios such as Fe/O that also depend upon the shock geometry (but see also Giacalone (2005) regarding turbulence at quasi-perpendicular shock waves). A source -temperature study of gradual SEP events found that 24 % of them originated at 2-4 MK like impulsive events but 69 % had source temperatures of $\leq 1.6$ MK, suggesting the ambient coronal plasma as a source (Reames, 2016).

As the SEPs from any source travel out to the spacecraft along the magnetic-field lines they may be scattered by Alfvén waves or similar magnetic turbulence. For small SEP events during quiet conditions, the transport may be scatter-free, meaning the scattering mean free path is comparable to the distance to the observer, normally $\approx 1$ AU (*e.g.* Mason *et al.*, 1989). Intensely streaming particles can generate or amplify existing turbulence so that the scattering mean free paths vary with time and space (*e.g.* Ng, Reames, and Tylka, 2003) eventually throttling and limiting the particle streaming (Reames and Ng, 1998, 2010). This wave generation also magnifies the scattering near the shock that makes possible acceleration to higher energies (*e.g.* Lee, 1983, 2005; Ng and Reames, 2008).

Recent studies of source temperatures (Reames, Cliver, and Kahler, 2014b, 2015; Reames 2016) involved selecting each temperature [$T$], which determined $Q$ and $A/Q$, fitting the observed abundance enhancements *vs.* $A/Q$, selecting the fit, and temperature, with the minimum value of $\chi^2$. This procedure determines not only $T$, but also the power of $A/Q$, which, according to simple diffusion theory, is expected to vary with time in a well-defined way. This theory, based upon power laws of element abundances, may well apply to small and moderate gradual SEP events. We select several events to test the theory, to explore possible values of the parameters, and to distinguish the effects of acceleration from those of transport. We then explore the patterns of correlations of abundances that are produced by the acceleration of plasma of different temperatures in gradual and impulsive SEP events.

The SEP abundances in this article were measured using the *Low Energy Matrix Telescope* (LEMT: von Rosenvinge *et al.*, 1995) onboard the *Wind* spacecraft, which measures the elements He through about Pb in the energy region from about 2 – 20 MeV amu$^{-1}$ with a geometry factor of 51 cm$^2$ sr, identifying and binning by energy interval the major elements from He to Fe onboard at a rate up to about 10$^4$ particles s$^{-1}$. Instrument resolution and onboard processing have been described elsewhere (Reames *et al.*, 1997; Reames, Ng, and Berdichevsky, 2001; Reames, 2000; Reames and Ng, 2004). Typical resolution of LEMT from He isotopes through Fe was shown by Reames *et al.* (1997) and by Reames (2014) and resolution of elements with $34 < Z < 82$ by Reames (2000, also 2015). The LEMT response was calibrated with accelerator beams of C, O, Fe, Ag, and Au before launch (von Rosenvinge *et al.*, 1995).

For this study we primarily use abundances in the 3 – 5 MeV amu$^{-1}$ interval, which is the lowest energy in LEMT where measurements of all species are available and intensities are greatest. This includes He, C, N, O, Ne, Mg, Si, S, and Fe. For Ar and Ca we use the 5 – 10 MeV amu$^{-1}$ interval which has better resolution (see Reames 2014). Species in the intervals of atomic number $34 \leq Z \leq 40$, $50 \leq Z \leq 56$, and $Z > 56$ were measured at 3 – 10 MeV amu$^{-1}$. For abundance enhancements, elements were normalized to O in the same energy interval and divided by the corresponding SEP coronal abundances given by Reames (2014).

The gradual SEP events considered in this article were originally selected by Reames (2016) and are listed with their properties in Appendix B therein. Impulsive SEP events considered are listed with their properties in the appendix in Reames, Cliver, and Kahler (2014a); their temperatures were determined by Reames, Cliver, and Kahler (2014b).

## 2. Diffusive Transport

If we assume that the scattering mean free path [$\lambda_X$] of species X depends upon as a power law on the particle magnetic rigidity [$P$] as $P^\alpha$ and upon distance from the Sun [$R$] as $R^\beta$ we can use the expression for the solution to the diffusion equation (Equation C3 in Ng, Reames, and Tylka 2003 based upon Parker 1963) to write the enhancement of element X relative to O as a function of time [$t$] as

$$X/O = L^{-3/(2-\beta)} \exp\{(1-1/L)\,\tau/t\}\; r^S \tag{1}$$

where $L = \lambda_X / \lambda_O = r^\alpha = ((A_x/Q_x)/(A_O/Q_O))^\alpha$ and $\tau = 3R^2/[\lambda_O(2-\beta)^2\,v]$ for particles of speed $v$. Since we compare ions at a fixed velocity, their rigidities are replaced by the corresponding values of $A/Q$. The factor $r^S$ represents any $A/Q$-dependent power-law enhancement at the source, prior to transport. For shock acceleration of impulsive suprathermal ions, it describes the power-law enhancement of the seed particles. For shock acceleration of the ambient coronal material, $S = 0$.

To simplify our processing, we can achieve a power-law approximation if we expand $\log x = (1-1/x) + (1-1/x)^2/2 + \ldots$ (for $x > \frac{1}{2}$). Using only the first term to replace $1-1/L$ with $\log L$ in Equation (1), we have

$$X/O \approx L^{\tau/t - 3/(2-\beta)}\, r^S \tag{2}$$

for $L > \frac{1}{2}$, as an expression for the power-law dependence of enhancements on $A/Q$ for species X.

More generally, we can write Equation (2) in the form $X/O = r^p$, where the exponent [$p$] is linear in the variable $1/t$, so that

$$p = a/t + b = \alpha\,\tau/t + S - 3\alpha/(2-\beta) \tag{3}$$

where we will see that $a$ and $b$ are directly measurable from the SEP-abundance observations.

## 3. Event Analysis

The analysis of each time period in each event follows that of Reames (2016). It is based on a theoretical knowledge of the temperature dependence of *A/Q* for each species and the power-law dependence of abundance enhancements on *A/Q*. Values of *Q vs. T* are given by Arnaud and Rothenflug (1985), Arnaud and Raymond (1992), and Mazzotta *et al.* (1998) up to Fe and by Post *et al.* (1977) above Fe and are shown for a range of *T* in Figure 1.

**Figure 1.** *A/Q* as a function of the theoretical equilibrium temperature for elements that are named along each curve. Points are spaced every 0.1 unit of log *T* from 5.7 to 6.8. Bands produced by closed electron shells with 0, 2, and 8 electrons are indicated, He having no electrons. Elements tend to move from one of these groups to another as the temperature changes.

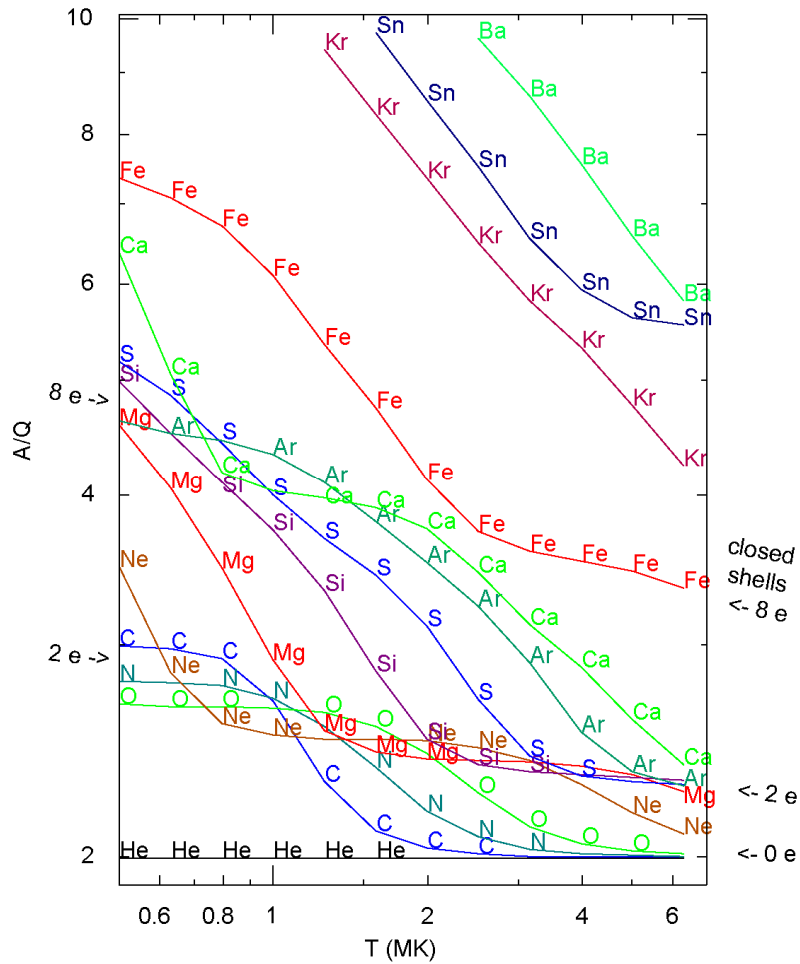

For each time period we calculate enhancements of the elements He, C, N, O, Ne, Mg, Si, S, Ar, Ca, and Fe, and the groups 34≤Z≤40, 50≤Z≤56, and Z>56, although the latter groups contribute little statistically. At a given *T*, the logarithmic pattern of abundance enhancements matches that of *A/Q*. The elements tend to group in *A/Q* around closed shells with zero, two, or eight orbital electrons. As *T* decreases, elements move from the group with no electrons to that with two electrons or and the group with two to the group with eight electrons. It is possible to estimate $T$ by observing the constituents of each group. However, we select the best $T$ from the quality of least-squares fits of log (enhancements) *vs.* log ($A/Q$) at each $T$.

The analysis follows as shown for the example SEP event in Figure 2 from Reames (2016). For each time period a temperature is selected which determines $Q$ and $A/Q$ for each particle species, the observed enhancements are least-squares fit as a function of $A/Q$ and the value of $\chi^2$ is noted. As other values of $T$ are selected, a plot of $\chi^2$ *vs.* $T$ is then made as shown in the upper-right panel of Figure 1 and the minimum $\chi^2$ gives the temperature and best-fit parameters for that time.

**Figure 2.** Clockwise from the lower-left panel are the intensities of H, C, and Fe during the 8 November 2000 SEP event, the enhancements in Fe and Ne during the event, the best-fit temperatures in color-coded 8-hour intervals, values of $\chi^2/m$ *vs.* $T$ for each time interval, and best-fit enhancements, relative to O *vs.* $A/Q$ and a sample least-squares fit for two 8-hour time intervals (Reames, 2016).

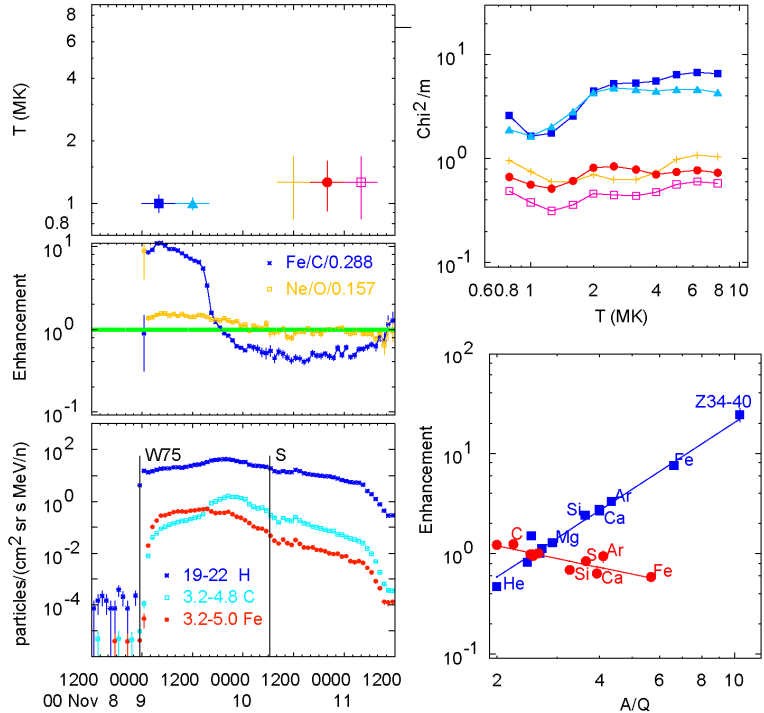

This procedure yields best-fit temperature, as a function of time, which was discussed by Reames (2016). It also yields a best-fit slope or power of $A/Q$ [$p$] as a function of time, which we show in the next section.

# 4. Parameters of Diffusive Transport

Figure 3 shows the evolution of $p$, the power of $A/Q$, described by Equation (3), for two gradual SEP events.

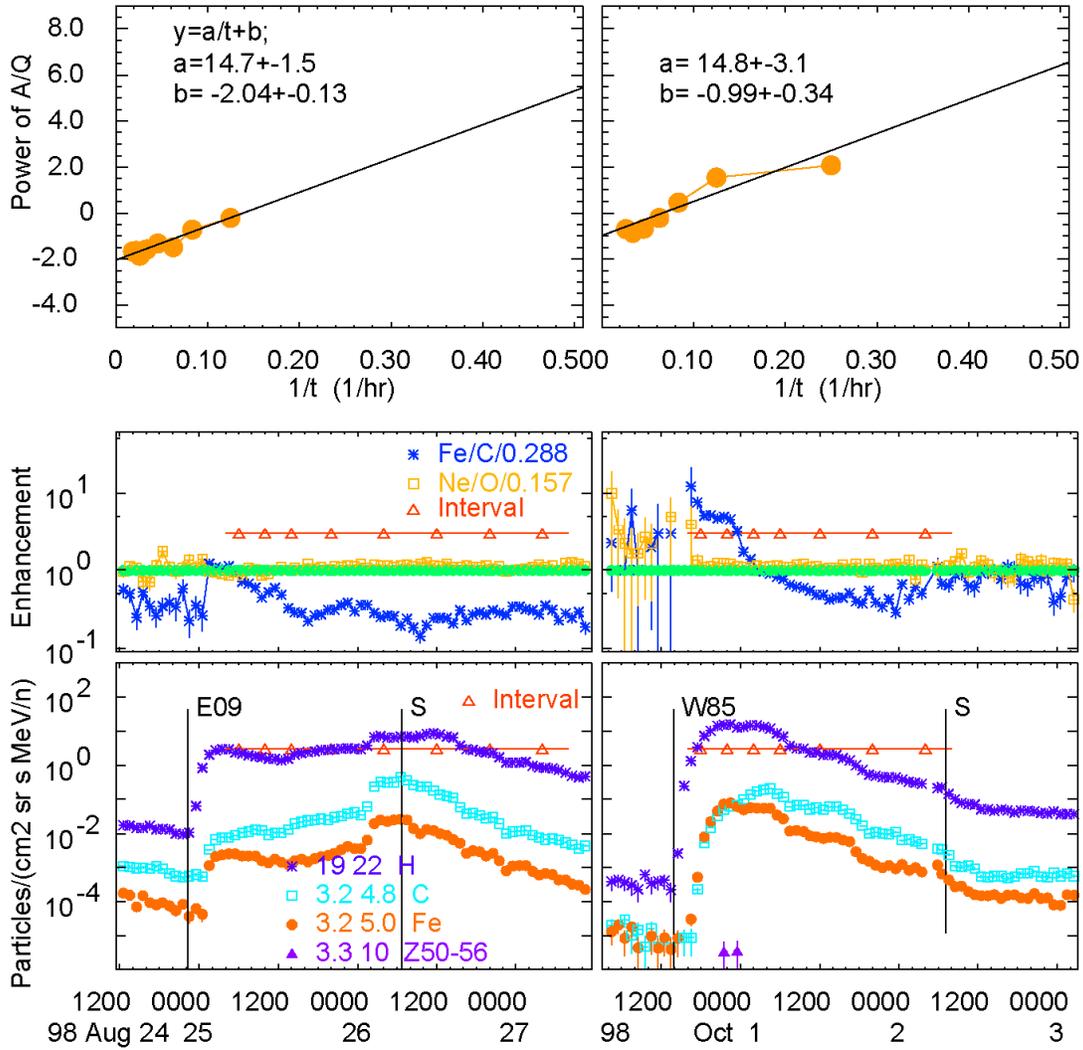

**Figure 3** Intensities *vs.* time in the lower panels, normalized enhancements *vs.* time in the middle panel, and *p* the power of *A/Q vs. 1/t*, the time from each flare onset, in the upper panel, for gradual SEP events of 24 August (left) and 30 September 1998 (right), with *T*= 1.6±0.2 and 1.3±0.2 MK respectively.

Figure 4 shows a similar analysis for two additional gradual SEP events that show considerable variation.

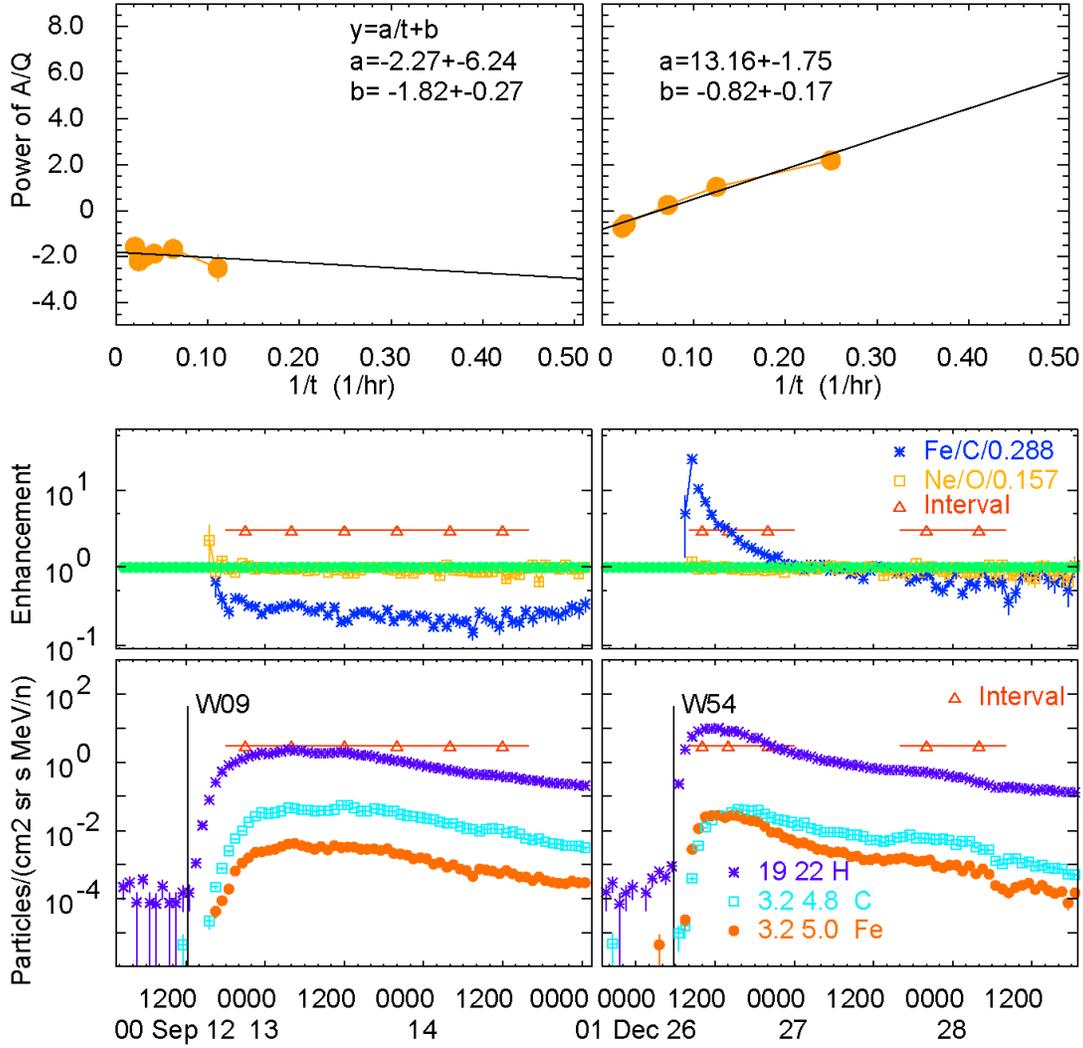

**Figure 4.** Intensities *vs.* time in the lower panels, normalized enhancements *vs.* time in the middle panel, and *p* the power of *A/Q vs.* 1/*t*, the time from each flare onset, in the upper panel, for gradual SEP events of 12 September 2000 (left) and 26 December 2001 (right), both with *T*= 1.6±0.2 MK.

The slope of *p* [*a*] depends inversely upon the scattering mean free path [$\lambda_O$] and can approach zero in a "scatter-free" condition when $\lambda_O$ is long, as it evidently is in the event on 12 September 2000. The four events considered so far have intercepts [*b*] ranging from -0.82±0.17 to -2.04±0.13 as shown in the figures. Since *b*= -3α / (2- β) when *S* =0, we have *α* between the Kolmogorov value of 0.33 and 0.66 for $\beta \approx 1$.

In Figure 5 we explore two SEP events of 6 November 1997 and 15 April 2001 with average source-plasma temperatures measured as 2.5±0.2 and 3.1±0.4 MK respectively (Reames, 2016). These events are expected to reaccelerate impulsive suprathermal ions with preexisting steep values of *S*.

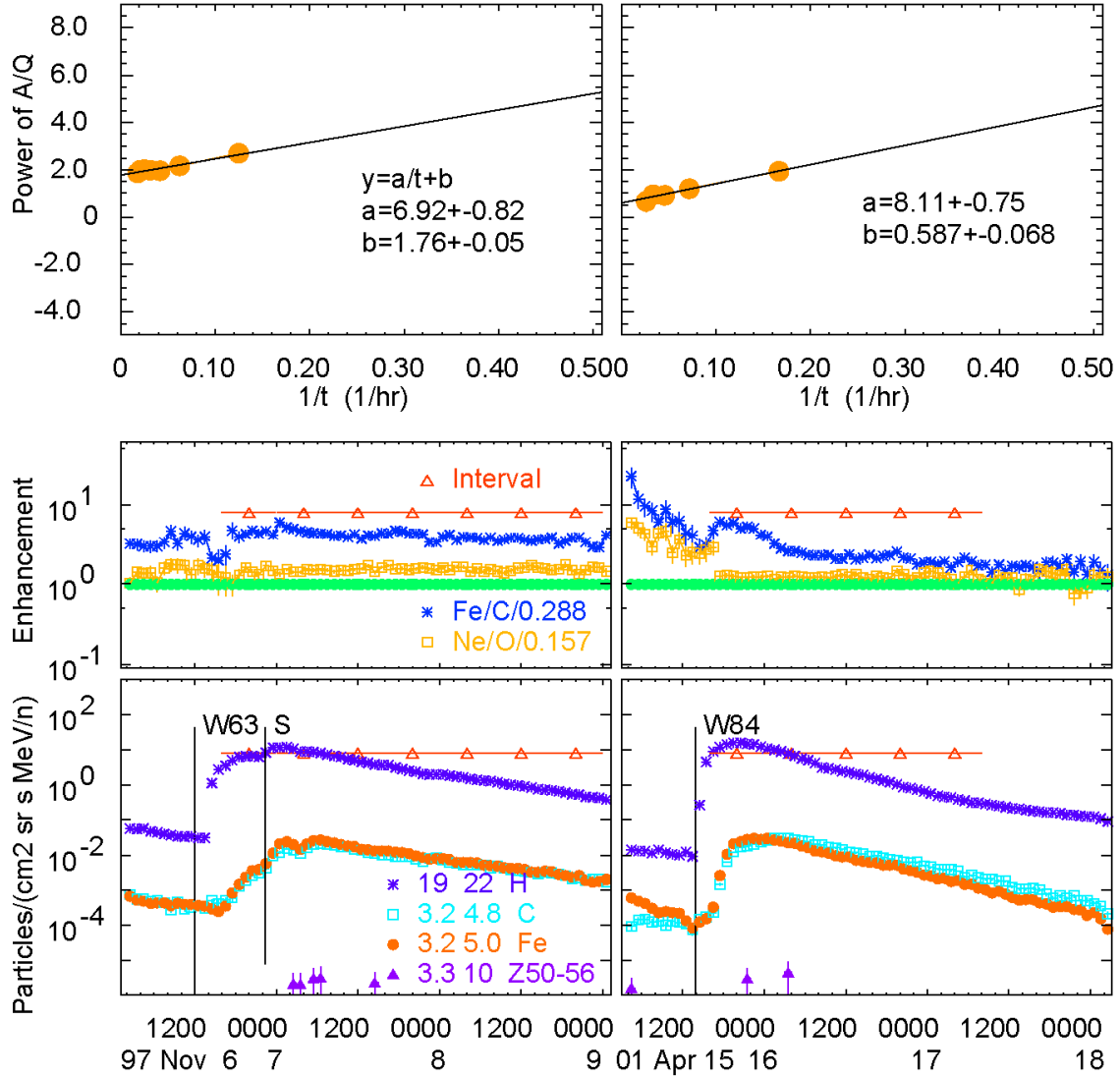

**Figure 5.** Intensities *vs.* time in the lower panels, normalized enhancements *vs.* time in the middle panels, and *p* the power of *A/Q vs.* 1/*t*, the time from each flare onset, in the upper panels, for gradual SEP events of 6 November 1997 (left) and 15 April 2001 (right). These events show source temperatures of 2.5±0.2 and 3.1±0.4 MK respectively indicating probable reacceleration of impulsive suprathermal ions.

If the value of $-3\alpha / (2-\beta)$ for these two events is -1 or -2, as for the previous events, then we have *S* between 2 and 4 for the injected impulsive ions. This is in the range discussed by Reames, Cliver, and Kahler (2014b) for impulsive SEP events. If we assume that $\alpha = 0.6$ and $\beta = 1$, values of slope $a = 15$ hour, imply $\lambda_O \approx 0.2$ AU while $a = 7$ hour implies $\lambda_O \approx 0.4$ AU. A scatter-free value of $\lambda_O \geq 1$ AU occurs for $a < 3$ hour. Many of the small impulsive SEP events with enhanced heavy elements are found to be scatter free (Mason *et al.*. 1989).

It is certainly not the purpose of this article to recommend the diffusion equation for all aspects of particle transport:

i) diffusion fails to explain event onsets (Tylka *et al.*, 2003; Reames, 2009a, b);

ii) in many small events, ions propagate scatter free (*e.g.*Mason *et al.*, 1989),

iii) streaming-generated waves in intense events cause complicated variations (Ng, Reames, and Tylka, 2003; Reames and Ng, 2010);

iv) the slow decline of event intensities late in events does not result from scattering but from magnetic trapping in a "reservoir" whose spectra decline adiabatically as the trapped volume increases with time; resevoirs may be scatter-free internally (*e.g.* review by Reames, 2013);

v) presumed lateral diffusion *via* the "birdcage model" misled the community for decades (*e.g.* review by Reames, 2015).

However, we do believe that the version of the parallel diffusion equation that is a power-law in abundances (Equations 2 and 3) fits the time dependence of a wide variety of gradual SEP events and provides a stronger basis for quantitative selection of the source plasma temperatures in gradual SEP events. In impulsive SEP events, acceleration produces a power-law dependence on $A/Q$ (Drake *et al.*, 2009; Reames, Cliver, and Kahler, 2014a, b) and scattering during transport contributes less (Mason *et al.*, 1989).

# 5. Patterns of Abundance

Given our new knowledge of the source plasma temperatures for 45 gradual SEP events from Reames (2016) and of the power of $A/Q$ for each eight-hour time interval within them, we now revisit the relationship between abundance ratios and the effect of these new parameters on previously enigmatic variations. Fe/O *vs.* C/He is shown in Figure 6 with temperature (lower panel) and power of $A/Q$ (upper panel) distinguished.

**Figure 6**. Each panel shows the normalized ratios of Fe/O *vs.* C/He for eight-hour periods during 45 gradual SEP events. The lower panel shows the circle size and color as the event source plasma temperature [$T$]. The upper panel shows the circle size as the power of $A/Q$ with the hotter (impulsive) events as open circles. Reference abundances are listed explicitly (*e.g.* He/O=57, C/O=0.42, and Fe/O = 0.131).

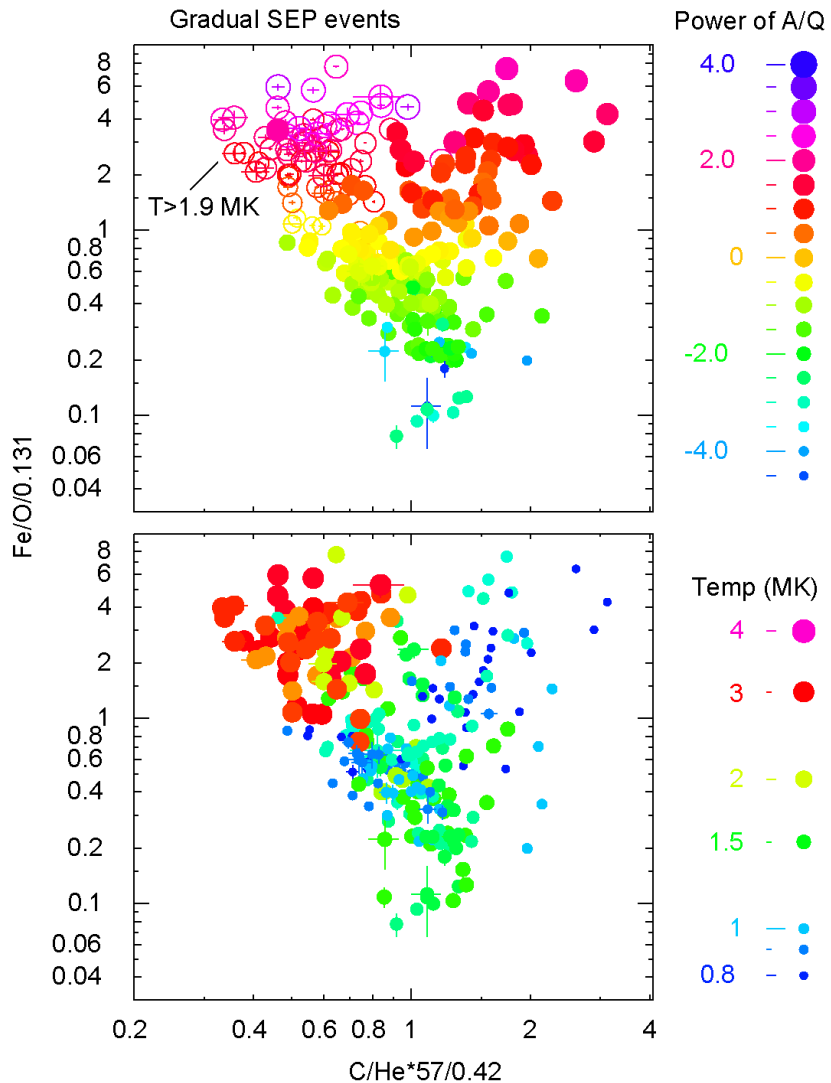

The large distribution of values of Fe/O *vs.* He/C in Figure 6 is clearly composed of regions of event with different $T$. The lower temperatures ($T \leq 1.9$ MK) show bands of different values of $p$ while the hotter (T > 1.9 MK) events, presumably from impulsive suprathermals, cluster in the upper left corner. Referring to Figure 1 we see that for $T \geq 2$ MK, C and He are both nearly fully ionized, have $A/Q = 2$, and they are thus indistinguishable to most acceleration and transport mechanisms. Yet they are not centered on a C/He136 value of unity. We explore this further in Figure 7.

**Figure 7**. Plots of normalized O/C *vs.* C/He for gradual SEP events are shown for three intervals of $T$ as indicated with the circle size and color varying with $p$, the power of $A/Q$. The direction of the expected decrease in $p$ with time is shown with arrows in the lower two panels.

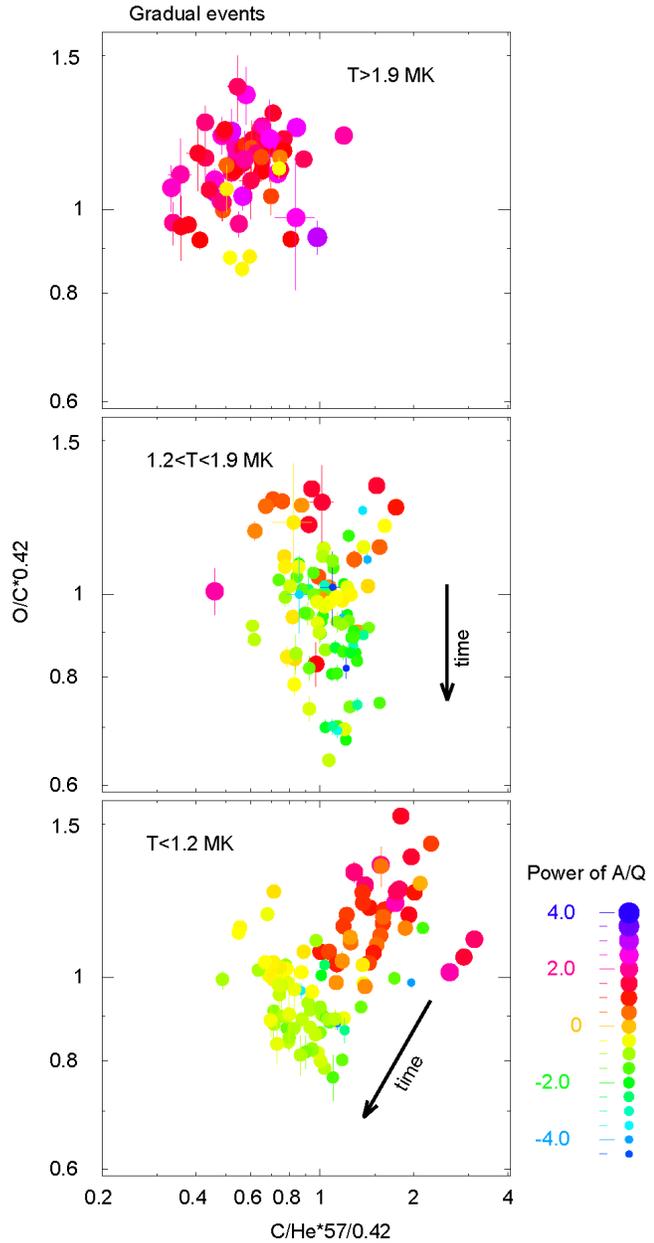

For the highest temperature interval in Figure 7 most of the enhancement in $p$ comes prior to shock acceleration and the He, C, and O ions are nearly fully stripped so there is little systematic variation of *p* with time. For the middle interval around $T \approx 1.5$ MK, O tends toward the band with two orbital electrons with $A/Q \approx 2.7$ while He and C have $A/Q \approx 2$, so O/C is enhanced early in events and declines, but C/He varies little. At the lower-temperature region near 1 MK, O is in the two-electron band while C is rising toward it and He is still stripped so both C and O are enhanced early, O more than C, and their enhancements decrease with time. Different-temperature plasma maps into different regions of enhancements.

The absolute value of C/He, when both are stripped, must give the reference abundance in the source plasma. This appears to be C/He136 ≈ 0.6 for $T > 1.9$ MK but C/He136 ≈ 1 seems reasonable for $T < 1.9$ MK. We will discuss this further below.

Figure 8 shows distribution of Ne/O and Si/Ne *vs.* Fe/C for events with different values of *T*. The events with $T \geq 2$ MK tend to be Fe-rich and Ne-rich like the impulsive events that may seed them (Reames, Cliver, and Kahler, 2014a, b).

**Figure 8**. Normalized ratios of abundances of Ne/O (lower panel) and Si/Ne (upper panel) are shown *vs.* Fe/C for gradual SEP events with the source plasma temperature [*T*] shown with the size and color of the circle.

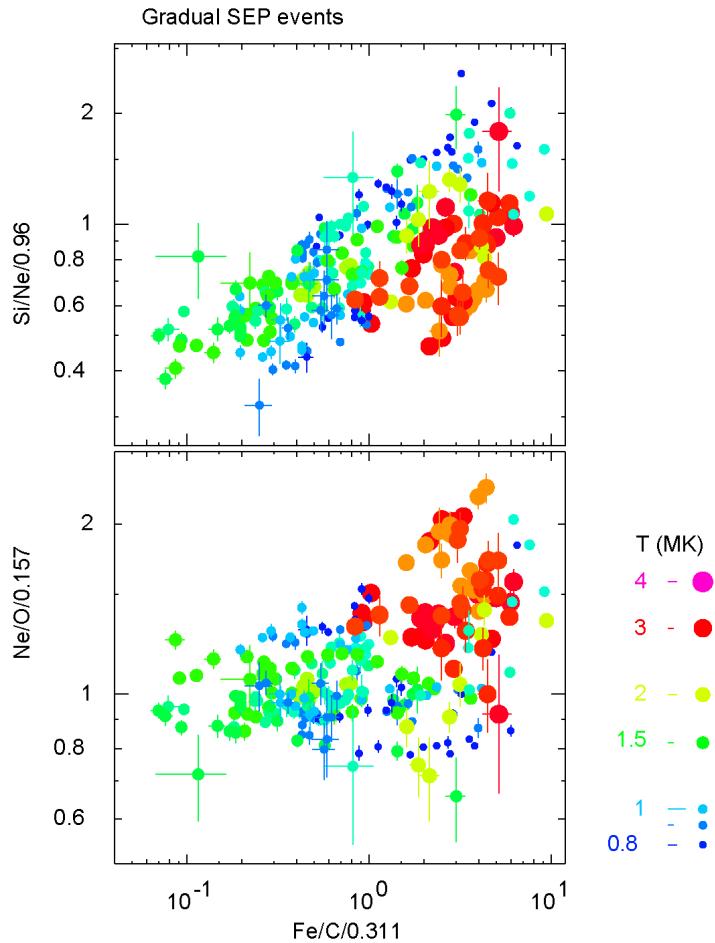

In Figure 9 we compare distributions of the enhancements of Fe/O *vs.* C/He for impulsive and gradual SEP events. The spread in C/He for impulsive events is large considering that C and He are both almost fully ionized above 2 MK. Some of the difference in the median values of C/He at 2.5 (orange) and 3.2 (red) MK for the impulsive events could result from the slight increase in *A/Q* for C. However, we cannot increase C/He, involving a few percent in *A/Q*, by an order of magnitude while only increasing Fe/O, involving a factor of ≈ 2 in *A/Q*, by an order of magnitude.

**Figure 9.** Distributions of enhancements of Fe/O *vs.* C/He are compared for impulsive and gradual events with symbol size and color indicating temperature.

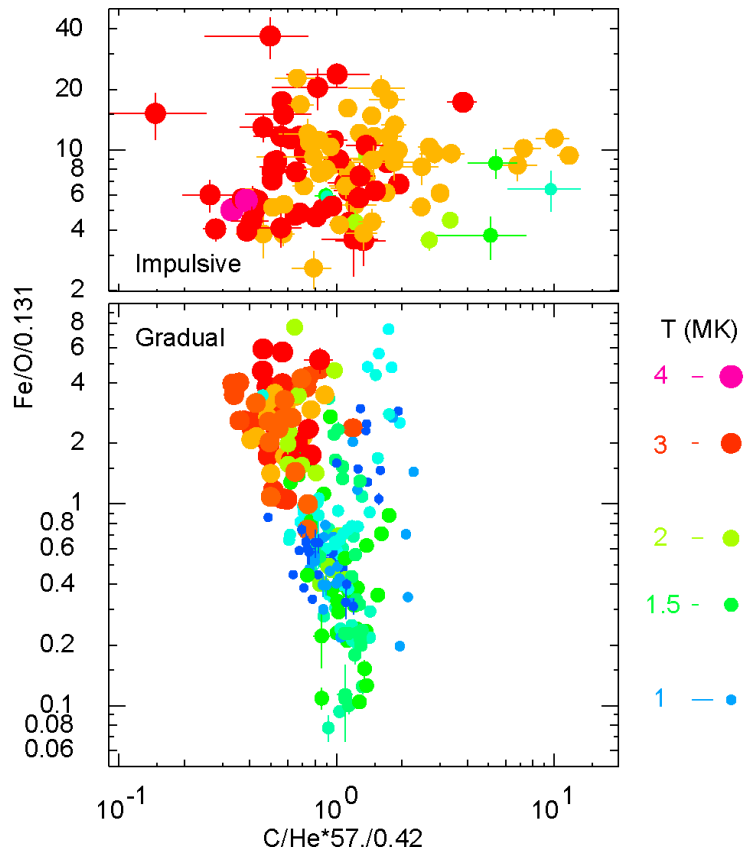

We explore the offset of the reference value of C/He for the impulsive events and the equally hot (> 1.9 MK) gradual events in Figure 10. The median value of C/He for the gradual events implies that the reference value of He/O should be 91 rather than 57 for these hot events. This value, shown by the dashed line, is not inconsistent with the hotter, *T* > 3 MK, impulsive events. The tight distribution of gradual events contrasts sharply with the broad distribution of the impulsive events in Figure 10 where both panels are plotted to the same scale.

**Figure 10.** Enhancements of O/C *vs.* C/He are compared, for gradual events with *T* > 1.9 MK and impulsive events with < 20 % errors. *T* is indicated by the size and color of the symbols. The median of the distribution of C/He for the gradual events, shown as a dashed line, implies a reference value for He/O of 91 rather than 57. Both panels are plotted on the same scale.

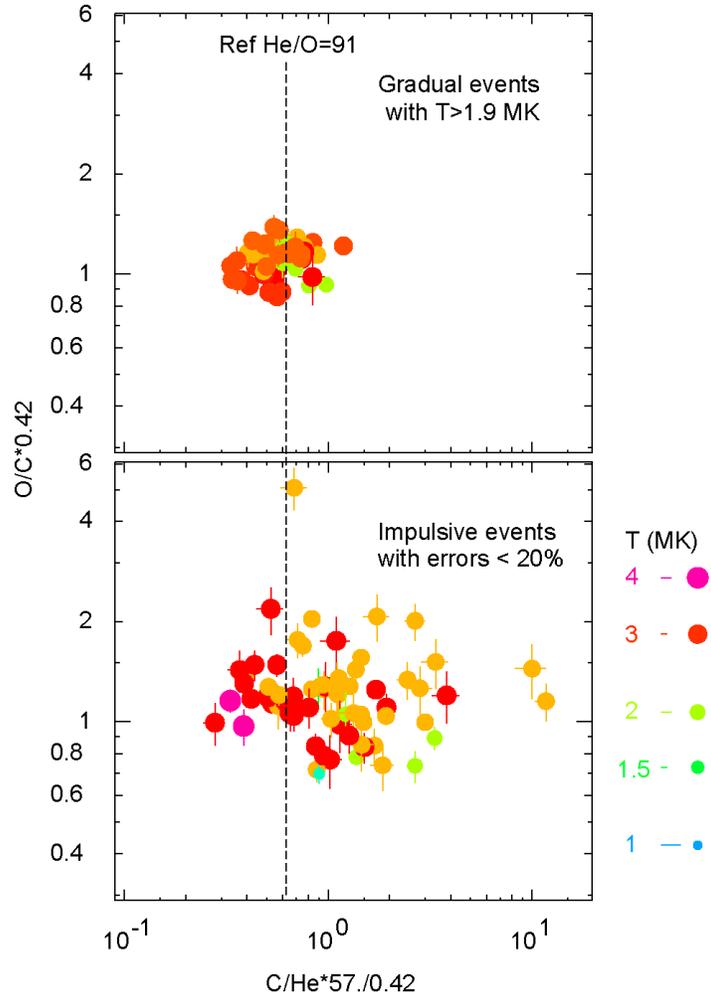

# 6. Discussion

For gradual SEP events, scattering of ions during interplanetary transport generates the power-law dependence of abundance enhancements or suppressions on *A/Q* of the ions that allow us to determine the source-plasma temperatures, which are also functions of *A/Q*. Diffusion theory supports this dependence and also predicts the time dependence of *1/t* that is commonly observed for gradual events. The temperature dependence is revealed by the way elements tend to group near different closed shells of zero, two, or eight orbital electrons as temperature changes, changing the pattern of *A/Q* and of the abundance enhancements.

### 6.1 Reference Abundances

Events with different temperatures occupy different regions in cross-plots of abundance ratios. Most interesting are the plots for the hottest plasma with $T > 1.9$ MK. In this region C is nearly fully ionized, like He, so both have $A/Q \approx 2$ and they cannot be distinguished by interplanetary electric or magnetic fields. The C/He ratio for these events suggests that the reference value of He/O should be 91 rather than 57. However, a reference He/O of 91 poorly describes the abundances in events with $T < 1.9$. Why He/O should differ between active regions and other regions of the ambient corona is not clear. No other differences in reference abundance are apparent.

A value of He/O = 91 is in better agreement with values measured in the solar wind (von Steiger *et al.*, 2000). However, using the solar-maximum solar-wind abundances from von Steiger *et al.* (2000) as a reference for the gradual SEP events systematically causes order-of-magnitude increases in the values of $\chi^2$ in the fits of enhancement *vs.* $A/Q$. Solar-wind abundances are a poor reference for the SEPs. Poor correlation between gradual SEP and solar-wind abundances has been noted previously (Mewaldt *et al.*, 2007; Desai *et al.*, 2006).

### 6.2 Impulsive SEPs vary more than Impulsive Suprathermal Ions

We find that an important difference between impulsive and gradual SEP events is the magnitude of event-to-event variation in abundances that should vary little, such as C/He discussed above (see Figure 10). Larger non-statistical variations among impulsive events are partly explained by the quantization of charge. Suppose, for example, that an event has 10 % $C^{+5}$ and 90% $C^{+6}$; the average value of $A/Q$ would be 2.033. An enhancement of $(A/Q)^5$ might be expected to enhance C/He by a factor of only 1.085. In reality, $C^{+5}$ would be enhanced by $(1.2)^5 = 2.49$ and $C^{+6}$ not at all, for a net enhancement of C/He of 1.15. A non-isothermal distribution in temperature involving a percentage of cooler plasma might also produce similar anomalous increases in C/He. The uncorrelated residuals of the least-squares fits and the effects of isotopes of Ne and Mg were discussed by Reames, Cliver, and Kahler (2014b). However, all of these effects do not easily explain an order-of-magnitude enhancement in C/He in a few events which is as large as the enhancement in Fe/O. Thus larger variations in impulsive SEP events are not explained by quantization, isotopes, temperature variations, or by the higher powers of $A/Q$ seen in these events. If impulsive suprathermal ions had the same variations as impulsive SEP ions, then those variations would also be reflected in gradual SEPs. Thus the gradual events cannot be produced by reacceleration of seed ions from a single impulsive event. *Impulsive SEPs have much larger variations than the impulsive suprathermals that apparently seed gradual SEP events.* Reaccelerated ions from a variable source would show comparable variations unless i) suprathermal seed ions are retained and thoroughly mixed or, ii) the suprathermal seed ions involve a broader spatial abundance average than the energetic impulsive SEP ions. The seed population of impulsive suprathermal ions may arise from multiple reconnection regions that are averaged by the shock, as we discuss in Section 6.4 below.

### 6.3 $^3$He-rich Impulsive Suprathermal Ions

Another measure of reacceleration of impulsive suprathermal ions by shocks is $^3$He/$^4$He. Observation of elevated $^3$He/$^4$He in large gradual events has long established that mechanism (Mason, Mazur, and Dwyer 1999), and the spectral variations of Fe/O (Tylka *et al.*, 2005; Tylka and Lee, 2006) support it. However, these involve small components of impulsive suprathermals, sometimes magnified by high-energy spectra. In our gradual events with $T \geq 2$ MK, the entire bulk of all energetic ions came from source plasma with $T \geq 2$ MK; if they should all come from impulsive suprathermal ions they would sometimes be very $^3$He rich.

At low intensities, LEMT can resolve $^3$He from $^4$He at ratios above 20 %, but resolution become difficult for LEMT at particle rates above about 1000 He ions sec$^{-1}$ because of spreading of the distributions in the increasing background. This is no problem in small impulsive events where all of the rates in the events studied by Reames, Cliver, and Kahler (2014b) were below 100 He-ions sec$^{-1}$. However, in our gradual SEP events rates can reach $10^5$ He ions sec$^{-1}$. Fortunately, nearly all events have some eight-hour periods with rates below 1000 He-ions sec$^{-1}$ and only one of these events, 11 April 2013, shows $^3$He/$^4$He above 20 %. This event has measured values of $^3$He/$^4$He of 35–40 % in three eight-hour periods, has $T = 2.0 \pm 0.3$ MK (Reames, 2016), and thus contains reaccelerated impulsive suprathermal ions. However, all other events are not $^3$He-rich, including the other 11 events in our sample with $T \geq 2$ MK. For the impulsive SEP events, about a third of the events have $^3$He/$^4$He above 20 % (Reames, Cliver, and Kahler, 2014b); combining several randomly selected events from this distribution would cause dilution of $^3$He/$^4$He in some cases.

Quite possibly, impulsive events sample small portions of a solar active region where abundances vary spatially, possibly involving magnetic loops of different ages, for example. In contrast, large shock waves may sample a much larger volume of the active region, averaging abundances over many smaller individual regions where reconnection produced suprathermal seed ions but not an energetic impulsive SEP event.

### 6.4 Nanojets?

It is not difficult to imagine a range of small regions of magnetic reconnection on open field lines in solar active regions that can accelerate ions, but not to MeV amu$^{-1}$ energies, being more numerous but less energetic than those jets that produce impulsive SEPs. Parker (1988) envisioned "nanoflares" with enough combined energy to heat the corona, a mechanism that is still important (*e.g.* Viall and Klimchuk, 2013) although nanoflares, alone, may not heat the corona (Klimchuk and Bradshaw, 2014). We require much much less energy to produce seed particles with typical enhancements of impulsive SEPs from jets that are smaller but more frequent. We prefer the term "jets", involving reconnection on *open* field lines, rather than "flares" where heating occurs as reconnection energy is trapped on closed loops (see Kahler, Reames, and Cliver, 2015). When a shock wave strikes the base of an observer's field line that connects to such an active region it accelerates ions that represent an average over the population of suprathermal ions seeded by many of these "nanojets" that were recently in progress. All share the source plasma temperature of the active region: 2 – 4 MK. This model for the seed population of gradual events with 2 – 4 MK source temperatures could produce the consistent average abundances seen in the upper panel of Figure 10. Mixing five to ten events with the variations of impulsive SEPs would calm the fluctuations to those of the gradual SEP events since $n$ independent samples would reduce the error in the mean by a factor of $n^{-1/2}$. Averaging nine nanojets would reduce the fluctuations in the gradual-event seed population from 30 % to 10 % . Shocks might also include ambient plasma from the active region that has the same temperature as the impulsive suprathermals but no enhancements of heavy elements or $^3$He, diluting the enhancements.

Mewaldt *et al.* (2003) examined the availability of Fe and $^3$He for the seed population during quiet and active solar periods. Since our model requires the presence of an active region for the shock to sample, there seems to be no conflict.

Nanojets might also contribute to the periods of persistent $^3$He seen by Wiedenbeck *et al.* (2008), of long-lived and recurrent sources (Bučík *et al.*, 2014, 2015) and, of course, to the substantial $^3$He abundances below 1 MeV amu$^{-1}$ in the seed population directly observed at 1 AU before and after the passage of shock waves (*e.g.* Desai *et al.*, 2003). Jets are also seen in solar-polar regions and, in principle, could contribute to gradual SEP events with cooler source plasma. However, there are no known SEP events with photospheric-like abundances, *i.e.* those lacking a strong dependence on the first ionization potential (FIP) of the elements, as seen in the fast solar wind (*e.g.* von Steiger *et al.*, 2000). SEPs show a fairly strong dependence on FIP (Reames, 1995, 2014, 2015) in the average abundances that serve as a reference for all of the events we study.

If multiple nanojets contribute impulsive suprathermal ions to the shock acceleration in gradual events with high source plasma temperatures, then these events should begin when the footpoints of the magnetic connection to the observer lies in or near an active region where nanojets abound. This agrees with the observation by Ko *et al.* (2013) that Fe-rich gradual SEP events are likely to be connected to solar active regions.

# 7. Summary

We find the following:

i) The equations of diffusive interplanetary transport may be cast into a form where
abundance enhancements are approximately power laws in *A/Q*.

ii) The time dependence of power-law enhancements in many gradual SEP events follows
the expected a linear power in time$^{-1}$ with scattering mean free paths $\geq$ 0.2 AU.
Many impulsive SEP events are scatter free.

iii) Source plasma temperatures may be deduced from the pattern of abundance enhancements *vs. A/Q* (Reames 2016).

iv) In cross-plots of abundances that are commonly studied, *e.g.* Fe/O *vs.* C/He, measurements from events of different temperatures or different times occupy different regions and traverse the plots in ways that can now be understood.

v) There is evidence that the reference abundance of He/O is higher (He/O=91) for events with $T \geq 2$ MK than for those with $T < 2$ MK (He/O=57). Non-SEP reference abundances, such as those from the solar wind, produce unacceptably poor fits.

vi) Non-thermal variations in abundances of impulsive SEP events are much larger than those in impulsive suprathermal seed ions accelerated by the shock wave in gradual events suggesting that the latter have been averaged over space or time as from a larger number of smaller reconnection regions (*i.e.* "nanojets"). More numerous nanojets might be a more reliable supply of the persistent enhancements of $^3$He that are commonly found in suprathermal ions (Desai *et al.*, 2003; Wiedenbeck *et al.*, 2008) than are the rarer but larger impulsive SEP events.

vii) Fe-rich gradual SEP events are often magnetically connected to solar active regions where frequent nanojets could supply impulsive suprathermal ions (Ko *et al.*, 2013).

viii) Incorporation of large enhancements of $^3$He/$^4$He > 20 % in impulsive suprathermals is lower (≈8 %) than in impulsive SEP events (≈30 %) perhaps because of averaging of suprathermals with ambient active-region plasma.

**Acknowledgments**: The author thanks Steve Kahler and Chee Ng for many helpful discussions and for valuable comments on this manuscript.

## Disclosure of Potential Conflicts of Interest

The author declares he has no conflicts of interest.

## References

Arnaud, M., Raymond, J.:1992, *Astrophys. J.* **398**, 394.

Arnaud, M., Rothenflug, R.:1985, *Astron. Astrophys. Suppl.* **60**, 425.

Breneman, H.H., Stone, E.C.: 1985, *Astrophys. J. Lett.* **299**, L57.

Bučík, R., Innes, D.E., Mall, U., Korth, A., Mason, G.M., Gómez-Herrero, R.: 2014 *Astrophys. J.* **786**,71.

Bučík, R., Innes, D.E., Chen, N.H., Mason, G.M., Gómez-Herrero, R., Wiedenbeck, M.E.: 2015 *J. Phys. Conf. Ser.* **642**, 012002.

Cliver, E.W., Ling, A.G.: 2007, *Astrophys. J.* **658**, 1349.

Cliver, E.W., Ling, A.G.: 2009, *Astrophys. J.* **690**, 598.

Cliver, E.W., Kahler, S.W., Reames, D.V.: 2004, *Astrophys. J.* **605**, 902.

Cohen, C.M.S., Mewaldt, R.A., Leske, R.A., Cummings, A.C., Stone, E.C., Wiedenbeck, M.E., Christian, E.R., von Rosenvinge, T.T. 1999 *Geophys. Res. Lett.* **26**, 2697.

Desai, M.I., Mason, G.M., Dwyer, J.R., Mazur, J.E., Gold, R.E., Krimigis, S.M., Smith, C.W., Skoug, R.M.: 2003, *Astrophys. J.* **588**, 1149.

Desai, M.I., Mason, G.M., Wiedenbeck, M.E., Cohen, C.M.S., Mazur, J.E., Dwyer, J.R., Gold, R.E., Krimigis, S.M., Hu, Q., Smith, C.W., Skoug, R.M.: 2004, *Astrophys. J.* **661**, 1156.

Desai, M.I., Mason, G.M., Gold, R.E., Krimigis, S.M., Cohen, C.M.S., Mewaldt, R.A., Mazur, J.E., Dwyer, J.R.: 2006, *Astrophys. J.* **649**, 470.

Drake, J.F., Cassak, P.A., Shay, M.A., Swisdak, M., Quataert, E.: 2009, *Astrophys. J. Lett.* **700**, L16.

Giacalone, J.: 2005, *Astrophys. J.* **624**, 765.

Gosling, J.T.: 1993, *J. Geophys. Res.* **98**, 18937.

Gopalswamy, N., Yashiro, S., Krucker, S., Stenborg, G., Howard, R.A.: 2004, *J. Geophys. Res.,* **109**, A12105.

Gopalswamy, N., Xie, H., Yashiro, S., Akiyama, S., Mälekä, P., Usoskin, I.G.: 2012 *Space Sci. Rev.* **171**, 23, DOI 10.1007/s11214-012-9890-4

Hsieh, K.C., Simpson, J.A.: 1970, *Astrophys., J. Lett.* **162**, L191.

Kahler, S.W.: 1992, *Annu. Rev. Astron. Astrophys.* **30**, 113.

Kahler, S.W.: 1994, *Astrophys. J.* **428**, 837.

Kahler, S.W.: 2001, *J. Geophys. Res.* **106**, 20947.

Kahler, S.W.: 2007 *Space Sci. Rev.* **129**, 359.

Kahler, S.W., Reames, D.V., Cliver, E.W.:2015, *34th Intl. Cosmic Ray Conf.,* in press (arXiv: 1509.09260)

Kahler, S.W., Cliver, E.W., Tylka, A.J., Dietrich, W.F.: 2012 *Space Sci. Rev.* **171**, 121.

Kahler, S.W., Sheeley, N.R.,Jr., Howard, R.A., Koomen, M.J., Michels, D.J., McGuire R.E., von Rosenvinge, T.T., Reames, D.V.: 1984, *J. Geophys. Res.* **89**, 9683.

Klimchuk, J.A., Bradshaw, S.J.: 2014, *Astrophys. J.* **791**, 60.

Ko, Y.-K., Tylka, A.J., Ng, C.K., Wang, Y.-M, Dietrich, W.F.: 2013, *Astrophys. J.* **776**, 92.

Lee, M.A.: 1983, *J. Geophys. Res.,* **88**, 6109.

Lee, M.A.: 1997, In: Crooker, N., Jocelyn, J.A., Feynman, J. (eds.) *Coronal Mass Ejections, Geophys. Monograph* **99**, AGU, Washington, 227.

Lee, M.A.: 2005, *Astrophys. J. Suppl.,* **158**, 38.

Lee, M.A., Mewaldt, R.A., Giacalone, J.: 2012, *Space Sci. Rev.,* **173**, 247.

Mazzotta, P., Mazzitelli, G., Colafrancesco, S., and Vittorio, N.: 1998 *Astron. Astrophys Suppl.* **133**, 403.

Mason, G.M.: 2007 *Space Sci. Rev.* **130**, 231.

Mason, G.M., Gloeckler, G., and Hovestadt, D.: 1984, *Astrophys. J.* **280**, 902.

Mason, G.M., Mazur, J.E., Dwyer, J.R.: 1999, *Astrophys. J. Lett.* **525**, L133.

Mason, G.M., Reames, D.V., Klecker, B., Hovestadt, D., von Rosenvinge, T.T.: 1986, *Astrophys. J.* **303**, 849.

Mason, G.M., Ng, C.K., Klecker, B., and Green, G.: 1989, *Astrophys. J.* **339**, 529.

Mason, G.M., Mazur, J.E., Dwyer, J.R., Jokipii, J. R., Gold, R. E., Krimigis, S. M.: 2004, *Astrophys. J.* **606**, 555.

Mewaldt, R.A., Cohen, C.M.S., Mason, G.M., Desai, M.I., Leske, R.A., Mazur, J.E., Stone, E.C., von Rosenvinge, T.T., Wiedenbeck, M.E.: 2003, *Proc. 28th Int. Cosmic Ray Conf.,* (Tsukuba) **6** 3229.

Mewaldt, R.A., Cohen, C.M.S., Mason, G.M., Cummings, A.C., Desai, M.I., Leske, R.A., Raines, J., Stone, E.C., Wiedenbeck, M.E., von Rosenvinge, T.T., Zurbuchen, T.H.: 2007, *Space Sci. Rev.* **130**, 207.

Meyer, J. P.: 1985, *Astrophys. J. Suppl.* **57**, 151.

Ng, C.K., Reames, D.V.: 2008 *Astrophys. J. Lett.* **686**, L123.

Ng, C.K., Reames, D.V., Tylka, A.J.: 2003, *Astrophys. J.* **591**, 461.

Parker, E.N.: 1963, *Interplanetary Dynamical Processes* Wiley, New York

Parker, E. N.: 1988, *Astrophys. J.* **330**, 474.

Post, D.E., Jensen, R.V., Tarter, C.B., Grasberger, W.H., Lokke, W.A.: 1977, *At. Data Nucl. Data Tables*, **20**, 397

Reames, D.V.: 1995, *Adv. Space Res.* **15** (7), 41.

Reames, D.V.: 1999, *Space Sci. Rev.,* **90**, 413.

Reames, D.V.: 2000, *Astrophys. J. Lett.* **540**, L111.

Reames, D. V.: 2009a, *Astrophys. J.* **693**, 812.

Reames, D.V.: 2009b, *Astrophys. J.* 706, 844.

Reames, D.V.: 2013, *Space Sci. Rev.* **175**, 53.

Reames, D.V.:2014, *Solar Phys.* **289**, 977, DOI: 10.1007/s11207-013-0350-4

Reames, D.V.: 2015, *Space Sci. Rev.,* **194**: 303, DOI: 10.1007/s11214-015-0210-7. arXiv: 1510.03449

Reames, D.V.: 2016, *Solar Phys.*, **291** 911, DOI: 10.1007/s11207-016-0854-9, arXiv: 1509.08948

Reames, D.V., Ng, C.K.: 1998, *Astrophys. J.* **504**, 1002.

Reames, D.V., Ng, C.K.: 2004, *Astrophys. J.* **610**, 510.

Reames, D.V., Ng, C.K.: 2010, *Astrophys. J.* **723**, 1286.

Reames, D.V., Stone, R.G.: 1986, *Astrophys. J.* **308**, 902.

Reames, D.V., Cliver, E.W., Kahler, S.W.: 2014a, *Solar Phys.* **289**, 3817, DOI: 10.1007/s11207-014-0547-1 (arXiv: 1404.3322)

Reames, D.V., Cliver, E.W., Kahler, S.W.: 2014b, *Solar Phys.* **289**, 4675, DOI: 10.1007/s11207-014-0589-4 (arXiv:1407.7838)

Reames, D.V., Cliver, E.W., Kahler, S.W.: 2015, *Solar Phys.* **290**, 1761, DOI: 10.1007/s11207-015-0711-2 (arXiv: 1505.02741).

Reames, D. V., Meyer, J. P., von Rosenvinge, T. T.: 1994, *Astrophys. J. Suppl.*, **90**, 649.

Reames, D.V., Ng, C.K., Berdichevsky, D.: 2001, *Astrophys. J.* **550**, 1064.

Reames, D.V., Ng, C.K., Berdichevsky, D.: 2001, *Astrophys. J.* **550**, 1064.

Reames, D. V., von Rosenvinge, T. T., and Lin, R. P.: 1985, *Astrophys. J.* **292**, 716.

Reames, D.V., Barbier, L.M., von Rosenvinge, T.T., Mason, G.M., Mazur, J.E., Dwyer, J.R.: 1997, *Astrophys. J.* **483**, 515.

Roth, I., Temerin, M.: 1997, *Astrophys. J.* **477**, 940.

Rouillard, A.C., Odstrčil, D., Sheeley, N.R. Jr., Tylka, A.J., Vourlidas, A., Mason, G., Wu, C.-C., Savani, N.P., Wood, B.E., Ng, C.K., *et al.*: 2011, *Astrophys. J.* **735**, 7.

Rouillard, A., Sheeley, N.R.Jr., Tylka, A., Vourlidas, A., Ng, C.K., Rakowski, C., Cohen, C.M.S., Mewaldt, R.A., Mason, G.M., Reames, D., et al.: 2012, *Astrophys. J.* **752**:44.

Serlemitsos, A.T., Balasubrahmanyan, V.K.: 1975 *Astrophys. J.* **198**, 195.

Shimojo, M., and Shibata, K.: 2000, *Astrophys. J.* **542**, 1100.

Temerin, M., Roth, I.: 1992, *Astrophys. J. Lett.* **391**, L105.

Tylka, A.J., Lee, M.A.: 2006, *Astrophys. J.* **646**, 1319.

Tylka, A.J., Cohen, C.M.S., Deitrich, W.F., Maclennan, C.G., McGuire, R.E., Ng, C.K., Reames, D.V.: 2001, *Astrophys. J. Lett.* **558**, L59.

Tylka, A.J., Cohen, C.M.S., Dietrich, W.F., Krucker, S., McGuire, R.E., Mewaldt, R.A., Ng, C.K., Reames, D.V, Share, G.H.: 2003, *Proc. 28th Int. Cosmic Ray Conf.,* (Tsukuba) **6** 3305.

Tylka, A.J., Cohen, C.M.S., Dietrich, W.F., Lee, M.A., Maclennan, C.G., Mewaldt, R.A., Ng, C.K., Reames, D.V.: 2005, *Astrophys. J.* **625**, 474.

Viall, N.M., Klimchuk, J.A.: 2013, *Astrophys. J.* **771**, 115.

von Steiger, R., Schwadron, N. A., Fisk, L. A., Geiss, J., Gloeckler, G., Hefti. S., Wilken, B., Wimmer-Schweingruber, R. F., Zurbuchen, T. H.: 2000, *J. Geophys, Res.* **105**, 27217.

von Rosenvinge, T.T., Barbier, L.M., Karsch, J., Liberman, R., Madden, M.P., Nolan, T., Reames, D.V., Ryan, L., Singh, S. *et al.*: 1995, *Space Sci. Rev.* **71**, 155.

Wang, L., Lin, R.P., Krucker, S., Mason, G.M.: 2012 *Astrophys. J.* **759**, 69.

Wiedenbeck, M.E, Cohen, C.M.S., Cummings, A.C., de Nolfo, G.A., Leske, R.A., Mewaldt, R.A., Stone, E.C., von Rosenvinge, T.T.: 2008, *Proc. 30th Int. Cosmic Ray Conf.* (Mérida) **1**, 91.